\documentclass[letter]{aa}

\usepackage{graphicx}
\usepackage{txfonts}
\usepackage{natbib}
\bibpunct{(}{)}{;}{a}{}{,} % to follow the A&A style.

\def\otres{[\ion{O}{iii}] $\lambda$ 5007 \AA\,\,}
\def\kms{km~s$^{-1}$}
\def\mone{$^{-1}$}
\def\mtwo{$^{-2}$}

\begin{document}

\title{Tracing jet--ISM interaction in young AGN:\\ correlations between \otres and 5-GHz emission}

\author{
A. Labiano\inst{1,2,3}
}
\offprints{Alvaro Labiano:\\ {\tt labiano at damir.iem.csic.es} \smallskip}

\institute{
Departamento de Astrof\'isica Molecular e Infrarroja, Instituto de Estructura de la Materia (CSIC), Madrid, Spain
\and
Kapteyn Astronomical Institute, Groningen, 9700 AV, The Netherlands %\\
\and
Department of Physics, Rochester Institute of Technology, Rochester, NY, 14623, USA%\\
}

\titlerunning{Interaction in young AGN: \otres and 5-GHz emission.}
\authorrunning{A. Labiano}

\date{ }

\abstract
{}
{To study the interaction between young AGN and their host galaxies based on their ionized gas and radio emission, and to analyze possible implications for the radio galaxy evolution.}
{The \otres line and 5-GHz radio properties are compared and studied on a large, representative sample of GPS and CSS (i.e., young) quasars and radio galaxies as well as large-scale sources using \otres line and 5-GHz radio data from literature and our observations. }
{Several correlations between the \otres line and 5-GHz radio emission have been found. The main result is that the \otres emission is strongly related to the GPS/CSS source size indicating that the \otres emission is clearly enhanced by the jet expansion through the host galaxy ISM. Shocks are the most likely enhancing mechanism, although jet-induced star formation could also be, partly, responsible for the \otres emission. The data also suggests a possible deceleration of the jet as it grows. In this case, however, the correlation is weak.}
{}

\maketitle

%________________________________________________________________

\section{Introduction}

We still know little about how radio galaxies are born and how they subsequently evolve but it is generally accepted that the GHz Peaked Spectrum (GPS) and Compact Steep Spectrum (CSS) radio sources are young, smaller versions of the large-scale powerful radio sources \citep[with a few exceptions, see e.g., ][]{Stanghellini05, Marecki06}.

The GPS and CSS sources are powerful but compact radio sources whose spectra are generally simple and convex with peaks near 1 GHz and 100 MHz respectively. The GPS sources are contained within the  extent of the optical narrow emission line region ($\lesssim 1$ kpc) while the CSS sources are contained within the host galaxy \citep[$\lesssim 15$ kpc, see ][ for a review]{O'Dea98}.

Current models of the evolution of powerful radio galaxies suggest that these sources propagate from the $\sim 10$ pc to Mpc scales at roughly constant velocity through an ambient medium whose density declines as $\rho(R) \propto R^{-2}$ while the sources decline in radio luminosity as $L_{rad} \propto R^{-0.5}$ \citep[e.g., ][ and references therein]{O'Dea02}. In this scenario, GPS sources would evolve into CSS sources and these into supergalactic-size sources\footnote{Recently, the High Frequency Peakers have been added to the sequence, as possible progenitors of GPS sources \citep[e.g., ][ and references therein]{Orienti07}.}. Such a scenario is consistent with the observed number densities of powerful radio sources as a function of linear size \citep[e.g.][]{O'Dea97,Fanti01}. However, to match observations, the radio jets of the young sources must slow down as they cross the host galaxy ISM and dim faster than predicted. The most likely mechanism to produce these effects is interaction of the radio source with the host environment \citep[see e.g.,][]{Young93,Carvalho94,Carvalho98}. 

The characteristics (size, radio power and young age) of GPS and CSS sources make them excellent probes of interaction (and evolution). Furthermore, they have not completely broken through the ISM, so these interactions are expected to be more important than in the larger sources. Observations of UV, \ion{H}{i} and, especially, of the ionized gas in GPS and CSS sources suggest the presence of such interactions \citep[e.g.,][]{Labiano08, Holt06, Labiano05, Axon00, Vries99, Vries97}.

Although evidence of interactions was found, previous investigations dealt with small samples or even just a few sources and, until now, there was no study of interaction of a large representative sample of GPS and CSS sources. Moreover, the sparseness of samples used did not allow for a general study of the consequences such interactions.

In order to examine the interaction between gas clouds and the radio source in a statistically significant manner, I collected a sample of almost one hundred sources, including GPS and CSS galaxies and quasars, as well as large-scale sources. I studied the properties of the \otres line and 5-GHz radio emission aiming to finding traces of interaction and the mechanisms responsible for it.

All calculations were made with H$_0$=71,  $\Omega_{\mathrm{M}}$ = 0.27, $\Omega_\Lambda = 0.73$, \citep{Spergel03}

\section{The sample}
\label{sec:sample}

I have compiled a representative sample of GPS and CSS as well as supergalactic-sized sources for which \otres line and 5-GHz observations were available. The sample consists of literature data \citep{O'Dea98, Gelderman94, Vries00b}, the 2-Jy radio sources \citep{Morganti97, Serego94, Morganti93, Tadhunter93}, STIS data \citep{O'Dea02, Labiano05} and new Kitt Peak observations of three GPS radio galaxies: \object{0554--026}, \object{0941--080}, \object{1345+125} (Table \ref{KPfits}). 

The Kitt Peak data consists of moderate dispersion ($\sim 200$ \kms) spectroscopy of the [\ion{O}{iii}] $\lambda$ 5007 \AA\ line obtained with the GoldCam CCD spectrograph at the Kitt Peak National Observatory 2.1-m telescope. I used a 600 line/mm grating with a 1.5" slit, covering the range between 5100 and 8100 \AA. The spectra were extracted and calibrated using standard IRAF procedures. After dark, bias, and flat-field correction, the images were background-subtracted to remove sky lines. Wavelength calibration was done using HeNeAr arcs taken after each exposure. The flux calibration and removal of atmospheric features were carried out using spectrophotometric standards from \cite{Massey88}.

The total number of sources included in the sample is 95 (21 GPS including our new observations, 22 CSS, and 52 large-scale sources).

%%%%%% TABLE 1 %%%%%%%
\begin{table}[t]
\begin{minipage}{\columnwidth}
\caption{ \otres measurements from Kitt Peak observations.}
\label{KPfits}
\centering
\begin{tabular}{cccc}
\hline
\hline
 Source       & Center & F$_{ [\ion{O}{iii}]}$ & FWHM \\
       		&	 \AA & 10$^{14}$ erg s\mone cm\mtwo & \kms \\
\hline
\object{0554--026}	& 6056.7 $\pm$ 0.7 & 1.49$\pm$ 0.13& 592 $\pm$ 66\\ 
\object{0941--080}	 & 6021.9 $\pm$ 0.5 & 1.09 $\pm$ 0.11 & 370 $\pm$ 33\\ 
\object{1345+125}	 & 5484.4 $\pm$ 0.6 & 4.51 $\pm$ 0.52&1187 $\pm$ 73 \\ 
\hline
\end{tabular}
\end{minipage}
\\ \\
\end{table}
%%%%%%%%%%%%%%%%

\section{Results and discussion}
\label{sec:results}

For the objects in the sample, the following properties of the \otres line and 5-GHz radio emission were compared: FWHM, luminosity, asymmetry, kurtosis (\otres line), as well as power, size and turnover frequency (radio emission).

\begin{figure}
\centering
\includegraphics[width=\columnwidth]{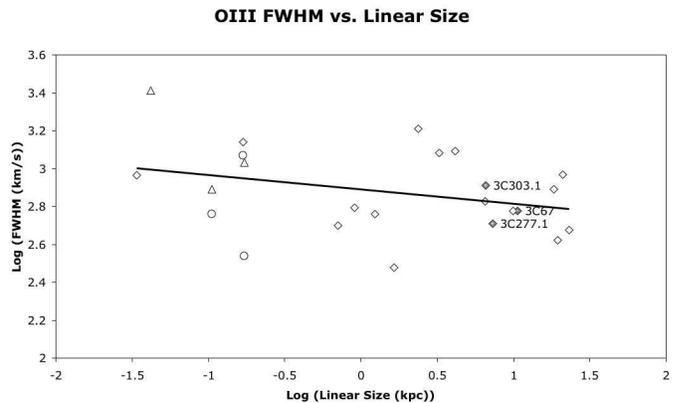}
\caption{Plot of the \otres FWHM of the sources in \cite{Gelderman94} (diamonds), \cite{Vries00b} (triangles) our Kitt Peak (circles) and STIS (shaded diamonds) observations, and a linear fit to the data. The ionization and kinematics of \object{3C~67}, \object{3C~277.1}, and \object{3C~303.1} were studied by \cite{Labiano05} and \cite{O'Dea02}.  \label{FWHMLS}}
\end{figure}

The \otres FWHM shows no correlation with radio power, turnover frequency, and \otres luminosity. Therefore, the shock velocity \citep[closely related to the FWHM, see e.g.,][]{Bicknell97} is independent of the radio source strength, and does not affect the luminosity of the ionized gas or the radio spectral properties of the source. However, a correlation between \otres FWHM and radio source size (Fig. \ref{FWHMLS}) could be present, suggesting a possible deceleration of the jet as it crosses the host galaxy:

\begin{center}
log FWHM $\simeq  2.89(\pm0.04) - 0.08(\pm0.05) \times$ log LS  
\end{center}

This correlation is not very strong but could be real given that the deceleration of the jet is required to explain observations \citep[e.g.,][]{O'Dea98} and the latest models predict deceleration in the jet (see e.g., Kawakatu \& Kuno 2008 in prep.). Unfortunately, there is not much data available on \otres FWHM to confirm any of such models.

%%%%%% FIG 4 %%%%%%%
%%%%%%%%%%%%%%%%

%%%%%% FIG 12 %%%%%%%
\begin{figure}
\centering
\includegraphics[width=\columnwidth]{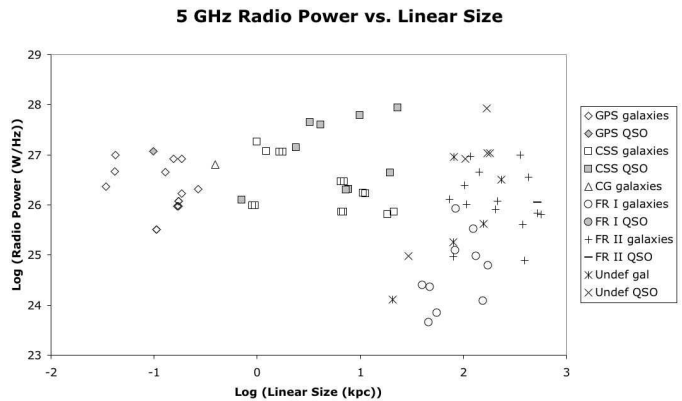}
\caption{Radio power at 5 GHz vs. linear size for GPS, CSS and large-scale radio sources. \label{RPLS}}
\end{figure}
%%%%%%%%%%%%%%%%

As expected \citep[see e.g.][Fig. \ref{RPLS}]{O'Dea97}, the sample also shows no correlation between radio power and radio size, showing that GPS, CSS and FR2\footnote{FRx stands for Fanaroff-Riley class x, \citep{Fanaroff74}.} sources are equally powerful (log P$_{\mathrm{5 GHz}}\sim10^{26-27}$ W/Hz) while FR1 sources tend to be fainter \citep[see e.g.,][]{Baum95, Zirbel95}. However, for giant sources ($\gtrsim 1$Mpc), the radio power seems to decrease with size \citep[e.g.,][]{Ishwara99}. 

It is also clear that quasars tend to be brighter in \otres than radio galaxies, consistent with the unification scenario: some of the \otres may be hidden by the torus in radio galaxies \citep[e.g.,][]{Hes93}. Also, the difference in \otres emission could be due to selection effects: quasars are usually found at higher redshifts. On the other hand, the sample could be missing fainter quasars with luminosities similar to radio galaxies. 

%%%%%% FIG 10 %%%%%%%
\begin{figure}
\centering
\includegraphics[width=\columnwidth]{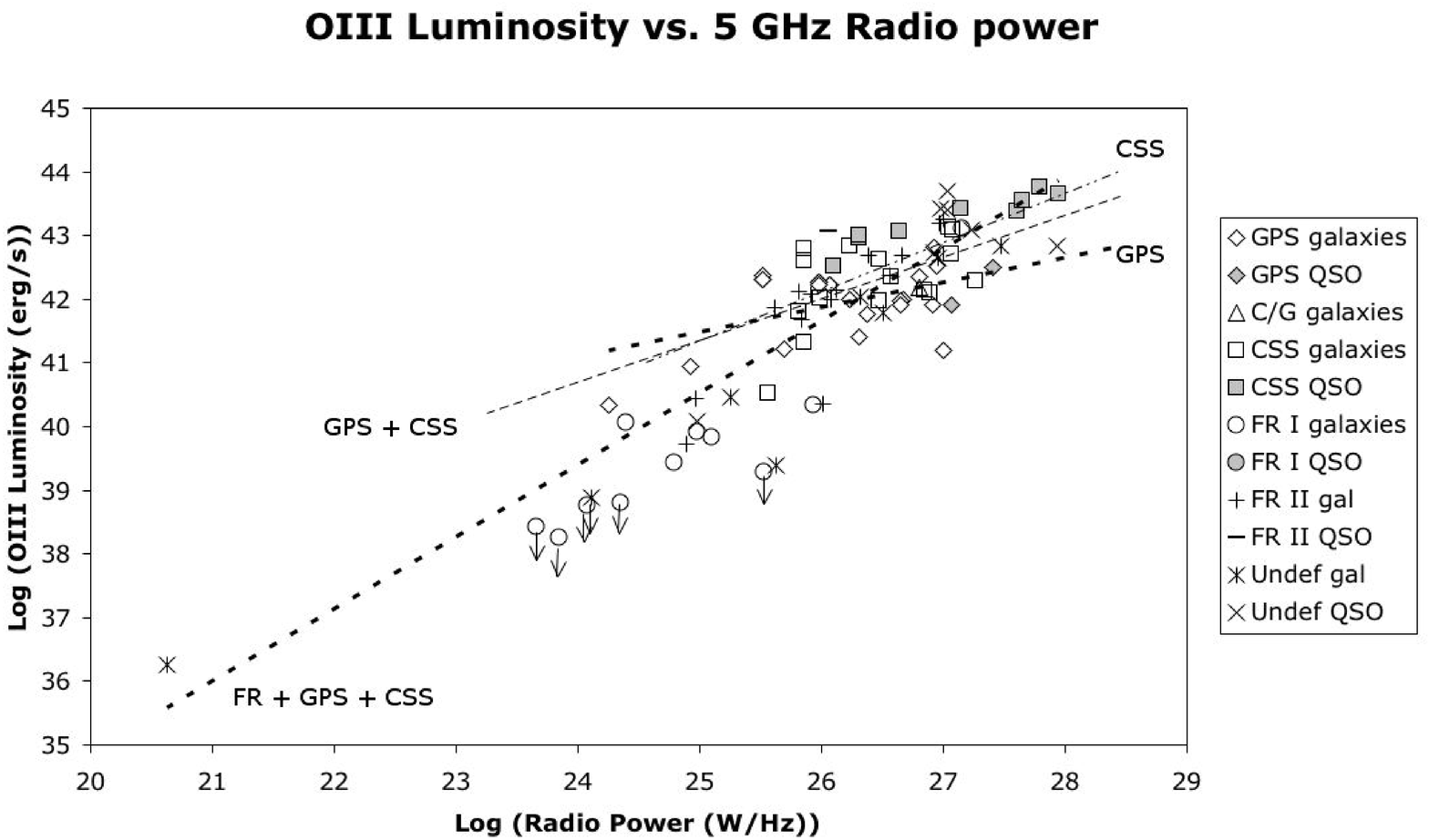}
\caption{Plot of the \otres luminosity vs. radio power at 5 GHz. Data for the CSS and GPS sources from \cite{Gelderman94,O'Dea98, Vries00b}, the 2-Jy sample and our Kitt Peak observations. Data for the large-scale radio sources from the 2-Jy sample. The IDs of the sources have been updated with \cite{Zirbel95}. I use "C/G" to name those sources with no clear ID as CSS or GPS. \label{O3RP}}
\end{figure}
%%%%%%%%%%%%%%%%

%%%%%% FIG 11 %%%%%%%
\begin{figure}
\centering
\includegraphics[width=\columnwidth]{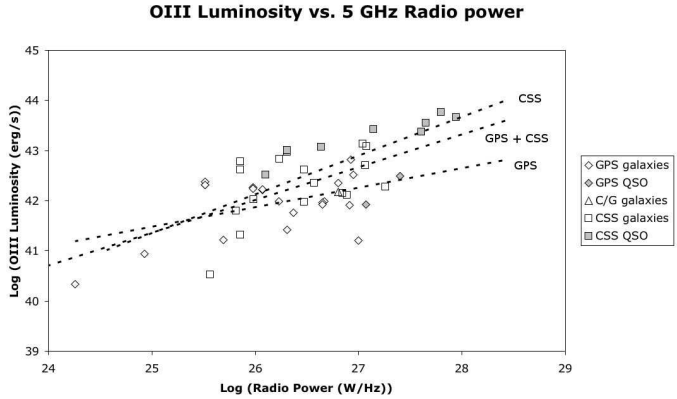}
\caption{As Fig. \ref{O3RP}, showing only the GPS and CSS sources. \label{O3RPzoom}}
\end{figure}
%%%%%%%%%%%%%%%%

The sample shows the known relation \citep[e.g.,][]{Baum89,Rawlings91} between \otres luminosity and radio power (Figs. \ref{O3RP} and \ref{O3RPzoom}) for powerful (log P$_{\mathrm{5 GHz}}\gtrsim10^{25}$ W/Hz) radio sources. This relation is usually explained as the AGN powering both the ionized gas and radio emission. The correlations for different groups of object in our sample are:

\begin{flushleft} GPS: \end{flushleft}
\begin{center}  log L$_{[\ion{O}{iii}]}$ = 32($\pm$4) + 0.4($\pm$0.1) $\times$ log P$_{\mathrm{5 GHz}}$ \end{center} 
CSS:
\begin{center}  log L$_{[\ion{O}{iii}]}$ = 22($\pm$4) + 0.8($\pm$0.2) $\times$ log P$_{\mathrm{5 GHz}}$\end{center}  
GPS + CSS: 
\begin{center} log L$_{[\ion{O}{iii}]}$ = 25($\pm$3) + 0.7($\pm$0.1) $\times$ log P$_{\mathrm{5 GHz}}$ \end{center} 
GPS + CSS + FR: 
\begin{center} log L$_{[\ion{O}{iii}]}$ = 12($\pm$2) + 1.13($\pm$0.07) $\times$ log P$_{\mathrm{5 GHz}}$ \end{center}

It should be noted that when combining NVSS and SDSS data (Best 2009, in preparation), this correlation is not present for fainter (log P$_{\mathrm{1.4 GHz}} \lesssim 24.5$ W/Hz) radio sources, suggesting that a different mechanism may be producing the radio and \otres emission in these sources.

There is no evident relation between kurtosis or asymmetry of the \otres line \citep[as defined by][]{Whittle85, Heckman81} with other properties of the line. Most of the sources have kurtosis lower than 1, suggesting the presence of broad wings in the gas. \cite{Whittle85} also finds the same for his sample of Seyfert galaxies. Most sources show asymmetry values close to zero in their \otres profile, suggesting the cocoon widens or narrows in a symmetric way  or there are no major variations between both sides. However, the ground spectra may lack sufficient resolution to discern more complex structures that could change the profile of the line.

\citet{O'Dea98} discovered that in the \citet{Gelderman94} sample, GPS galaxies tend to have lower \otres luminosity than CSS galaxies but it was not clear if the trend would be followed by a larger sample with different selection criteria, which was the case of quasars and large FR sources. Also, the implications of the existence of that trend on the radio source (and host galaxy) evolution are not clear\footnote{A direct consequence is that it is more difficult to find optical counterparts of GPS sources than CSS sources.}. To asses these issues, I looked for this trend in the sample presented here, which included not only GPS and CSS radio galaxies but also quasars and large-scale sources. I found that the GPS and CSS sources (galaxies and quasars) showed a strong correlation between \otres luminosity and size of the radio source\footnote{Note that the trend is also suggested by Fig. \ref{O3RPzoom} and the different correlations between radio power and \otres luminosity for GPS and CSS sources.}:

\begin{flushleft} GPS + CSS: \end{flushleft}
\begin{center} log L$_{[\ion{O}{iii}]}$ = 42.43($\pm$0.09) + 0.46($\pm$0.09) $\times$ log LS$_{\mathrm{5 GHz}}$ \end{center}
 
\begin{flushleft} GPS$^*$ + CSS:  \end{flushleft}
\begin{center} log L$_{[\ion{O}{iii}]}$ = 42.44($\pm$0.08) + 0.4($\pm$0.1) $\times$ log LS$_{\mathrm{5 GHz}}$ \end{center}

GPS$^*$ means the complete sample of GPS without the smallest source (\object{1718--649}), to test if it is dominating the correlation. 

In principle, this correlation could be due to the AGN enhancing both the radio and emission gas luminosities through photons. However, for the same radio power, small sources (GPS) are systematically fainter in \otres than larger (CSS) sources. 

\begin{figure}
\centering
\includegraphics[width=\columnwidth]{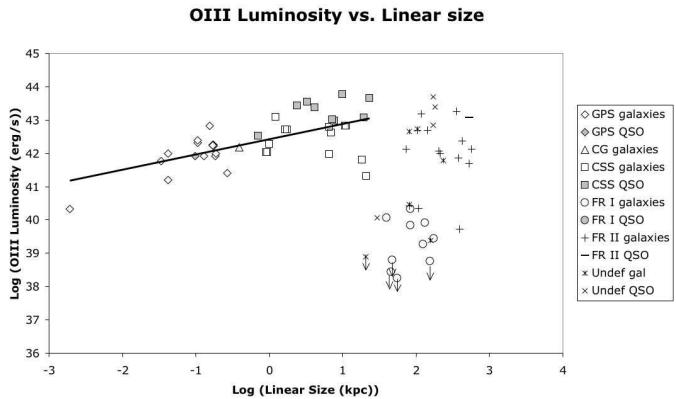}
\caption{Plot of the [\ion{O}{iii}] luminosity vs. linear size of the radio source, showing the correlation for GPS and CSS sources. Data from the same references as Fig. \ref{O3RP}. \label{O3LS}}
\end{figure}

I propose a scenario where the expansion of the radio source through the host ISM is triggering and/or enhancing the \otres emission through direct interaction. Some contribution from the AGN light must be present but AGN light alone would not produce a correlation with size. Furthermore, the fact that the correlation disappears for larger radio sources ($\gtrsim 15-20$ kpc) supports this model: once the radio lobes leave the host galaxy, the \otres luminosity drops (Fig. \ref{O3LS}). 

This scenario is also supported by previous observations providing evidence of strong interaction between the jet and surrounding ISM, as well as proof of shock-ionized \otres emission \citep[e.g.,][]{Labiano05, O'Dea03}. Jet--ISM interactions are also found through  \ion{H}{i} studies \citep[e.g.,][ and references therein]{Labiano06, Holt06} and predicted by jet expansion models \citep[e.g.,][]{Jeyakumar05, Saxton05}.

Another interesting enhancing mechanism to consider is that the jet could, at least partly, boost \otres emission through indirect mechanisms such as jet-induced star formation (Labiano et al. 2008, in prep.) but new data are needed to study the possible contribution of recently formed stars. The hosts of GPS and CSS sources are usually elliptical galaxies so it is unlikely that the average/normal stellar population of the host galaxy has a strong contribution to the \otres emission. These ``normal'' stars would, however, not create a correlation with radio jet size (and jet-induced stars would).

The correlations are as follows:\\

GPS:
\begin{center} log L$_{[\ion{O}{iii}]}$ = 42.8($\pm$0.2) + 0.8($\pm$0.2) $\times$ log LS$_{\mathrm{5 GHz}}$ \end{center}
 
GPS$^*$:
\begin{center} log L$_{[\ion{O}{iii}]}$ = 42.5($\pm$0.4) + 0.5($\pm$0.4) $\times$ log LS$_{\mathrm{5 GHz}}$ \end{center}

CSS:
\begin{center} log L$_{[\ion{O}{iii}]}$ = 42.6($\pm$0.2) + 0.2($\pm$0.3) $\times$ log LS$_{\mathrm{5 GHz}}$\end{center} 

The correlation tends to disappear when the sample is divided into different types of sources. This could be due to low statistics or to the smaller range of sizes covered by each type. GPS could show lower \otres emission due to high obscuration. However, it is more likely that young compact sources are too small to strongly affect their environment. This effect has also been observed in star formation histories of GPS hosts \citep{Labiano08}. The UV luminosities of GPS sources seem to be as high as those of CSS sources. Furthermore, the UV luminosity of GPS sources could be correlated with their radio power \citep{Labiano08}. These two effects suggest that obscuration is not too strong in GPS sources or, at least, similar to that in CSS.

%%%%%% FIG 13 %%%%%%
%%%%%%%%%%%%%%%%

Concerning the overall scenario of radio source evolution, where GPS and CSS sources evolve into the larger FR sources, the visual inspection of Fig. \ref{O3LS} suggests that CSS sources would evolve into FR2. Some authors also found a possible decreasing trend linking FR2 to FR1 (Best 2008, private communication) with increasing size. However, our sample may lack enough supergalactic-sized sources to thoroughly address evolution beyond $\sim15-20$ kpc and the FR2 -- FR1 connection is beyond the scope of this letter. Extensive discussions on the FR2 -- FR1 connection can be found in the literature: e.g., \citet{Muller04, Best05, Wold07} and references therein or classical papers such as \citet{Baum95, Zirbel95}.

\section{Summary of main results and future work}

The aim of this project was to improve our understanding of radio jet--host interaction in young AGN through studying the \otres line and 5-GHz radio emission properties of GPS and CSS sources. I compiled a large, representative sample of GPS and CSS quasars, combined with FR 1 and FR 2 sources (to help establish a evolution timeline) from published data, as well as our observations. 

The main result of the study is that the \otres emission is clearly enhanced by the jet expansion through the host galaxy ISM. This is consistent with previous observations as well as numerical models of jet expansion. However, further work is required (Labiano et al. 2008, in prep.):

\begin{itemize}
\item the supergalactic-size sample needs to be widen, to establish evolution when the radio lobes leave the host galaxy;
\item evaluate the strength of the jet shocks and study their capability to form stars (which could enhance the \otres) as well as the shocks direct contribution to the \otres emission;
\item study of star formation and AGN tracers (such as X-rays, 24 and 70 $\mu$m dust, line ratios, PAH, etc.) to identify and evaluate different mechanisms and contributions to the \otres emission;
\item apply and improve jet expansion models to reproduce the results.

\end{itemize}

A parallel result to the \otres emission -- radio source size correlation is that it will be more difficult to find optical counterparts of GPS than CSS sources.

The data suggests a possible deceleration in the jet as it crosses the host galaxy ISM, which is required by most radio source evolution models. However, the correlation is too weak and a much larger sample, with higher resolution spectra, is needed.

The sample also reflects some already known results such as quasars showing lower \otres luminosity than radio galaxies, radio power being correlated with  \otres luminosity, and radio power being independent of the size of the radio source.

\begin{acknowledgements}

This research has made use of the NASA/IPAC Extragalactic Database (NED) which is operated by the Jet Propulsion Laboratory, California Institute of Technology, under contract with the National Aeronautics and Space Administration. I would like to especially thank all the participants in the 4th GPS and CSS workshop for their extremely useful comments. Also, my gratitude to S. A. Baum, C.P. O'Dea and P. D. Barthel for their hospitality and many fruitful discussions. Last but not least, I would like to thank the referee, Dr. Andrzej Marecki, for thoroughly revising the manuscript and for plenty of useful comments which improved the letter.
\end{acknowledgements}

\bibliographystyle{aa}
\bibliography{../ALORefs}

\end{document}